\documentstyle[11pt,epsf]{article}
\newcommand{\al}{\alpha}
\newcommand{\ar}{{\cal A}}

\newcommand{\ba}{\begin{array}}

\newcommand{\bc}{\begin{center}}
\newcommand{\be}{\begin{equation}}
\newcommand{\ber}{\begin{eqnarray}}
\newcommand{\bt}{\beta}

\newcommand{\cl}{\clubsuit}

\newcommand{\de}{\delta}

\newcommand{\ea}{\end{array}}
\newcommand{\ear}{\end{eqnarray}}
\newcommand{\ec}{\end{center}}
\newcommand{\ee}{\end{equation}} 
\newcommand{\eei}{\end{equation}\indent\indent}
\newcommand{\el}{\ell}

\newcommand{\fr}{\frac}

\newcommand{\ga}{\gamma}

\newcommand{\Lg}{{\cal L}}
\newcommand{\la}{\lambda}

\newcommand{\na}{\nabla}

\newcommand{\p}{\partial}

\newcommand{\rh}{\rho}

\newcommand{\si}{\sigma}

\newcommand{\sq}{\sqrt}

\newcommand{\ta}{\tau}

\newcommand{\vb}{\verb}

\def\case#1/#2{\textstyle\frac{#1}{#2} }
\begin{document}
\title{The String Deviation Equation\\}
\author{Mark D. Roberts, \\\\\\
Department of Mathematics and Applied Mathematics, \\ 
University of Cape Town, South Africa\\\\\\
roberts@gmunu.mth.uct.ac.za} 
\date{\today}
\maketitle
\vspace{1.0truein}
\bc Eprint:  http://xxx.lanl.gov/abs/gr-qc/9810043  \ec
\bc Comments:  18 pages   57580 bytes, no diagrams,  LaTex2e.\ec
\bc 2 KEYPHRASES:\ec
\bc String Deviation:~~  Geodesic Deviation.\ec
\bc 1999 PACS Classification Scheme:\ec
\bc http://publish.aps.org/eprint/gateway/pacslist \ec
\bc 12.25+e\ec
\bc 1991 Mathematics Subject Classification:\ec
\bc http://www.ams.org/msc \ec
\bc 81T30.\ec
\newpage
\begin{abstract}
It is well known that the relative motion of many particles can be described
by the geodesic deviation equation.   Less well known is that the geodesic 
deviation equation can be derived from the second covariant variation of 
the point particle's action.   Here it is shown that the second covariant
variation of the string action leads to a string deviation equation.
This equation is a candidate for describing the relative motion 
of many strings,
and can be reduced to the geodesic deviation equation.
Like the geodesic deviation equation,  the string deviation 
equation is also expressible in terms of momenta and projecta.
It is also shown that a combined action exists, the first variation of 
which gives the deviation equations.   The combined actions allow the deviation
equations to be expressed soley in terms of the Riemann tensor,  
the coordinates,   and momenta.   In particular geodesic deviation can be 
expressed as:
\bc$\dot{\Pi}^{\mu}=R^{\mu}_{.\al \bt \ga}r^{\ga}P^{\bt}\dot{x}^{\al}$,\ec
and string deviation can be expressed as:
\bc$\dot{\Pi}_{\ta}^{\mu}+\Pi'^{\mu}_{\si}
=R^{\mu}_{. \al \bt \ga}
r^{\ga}(P_{\ta}^{\bt}\dot{x}^{\al}+P_{\si}^{\bt}x'^{\al})$.\ec
\end{abstract}
\newpage
\section{Introduction} \label{sec:intro}
The second covariant variation of the point particle action produces 
the geodesic deviation equation 
Synge (1926) \cite{bi:synge},  Hawking and Ellis (1973) \cite{bi:HE} and 
Bazanski (1977) \cite{bi:bazanski}.
Here the second variation method 
is applied to the string action Scherk (1974) \cite{bi:scherk} 
to produce a string deviation equation.
Second order variations have been applied to the metric in quadratic 
gravitational theories Barth and Christensen (1983) \cite{bi:BC}.
Geodesic deviation equations are useful in investigating the 
structure of cosmological models Ellis and Van Elst (1997) \cite{bi:EV},
are used to study effective photon mass in higher order theory,
Mohantz and Prasanna (1997) \cite{bi:MP},
are used to study the stability of Bianchi models,
DiBari and Cipriani (1998) \cite{bi:DBC},
and also are used to study the geometry of implusive gravitational waves,
Steinbauer (1998) \cite{bi:steinbauer}
and Kunzinger and Steinbauer (1998) \cite{bi:KS}.
The equations for geodesic deviation sometimes can be solved by
the inverse scattering method,  Varlamov (1998) \cite{bi:varlamov}.
The string deviation equations might have application in studying 
objects such as binary cosmic strings,  DasGupta and Rohm (1996) \cite{bi:DGR},
cosmic strings with wiggles,  Kim and Sikivie (1994) \cite{bi:KimS},
string interaction and collision,  Letelier et al (1993) \cite{bi:LGG},
and string evolution,  Ausin,  Copeland,  and Kibble (1993) \cite{bi:ACK}.
Also they might have application to the loop space approach to 
fundamental strings,  Bowick and Rajeev (1987) \cite {bi:BR}. 
The string deviation equation has one separation vector which 
connects adjacent strings.   Both deviation equations can be derived from 
combined actions as is shown here in section \ref{sec:combined}.   
At least five problems which are left for future investigation
include the following.  
{\it Firstly} the geometrical interpretation of the 
string deviation equation is not looked at in detail.
Reduction of the string deviation equation
gives the geodesic deviation equation.
Both deviation equations are expressible 
soley in terms of momenta and projecta.
These can be defined for both arbitary particle and string Lagrangians
and specifically for the standard case.
{\it Secondly} the production of higher order deviation equations 
corresponding to the third covariant variation,  fourth covariant 
variation and so on \ldots
{\it Thirdly} the construction of momenta and phase space with a view to 
quantizing the system by traditional canonical methods:   this has been done 
for the geodesic deviation equation Roberts (1996) \cite{bi:mdr1}.
{\it Fourthly} the implication for the algebra obtained by 
Fourier transforming string modes.   {\it Fifthly} the application to   
hypothetical physical strings such as cosmic strings and fundamental strings.
   
There are at least three ways of generalizing the point particles 
action.   The {\bf first} is by changing the number of dimensions of the 
ambient space or the number of of the intrinsic space,  changing the 
dimensions of the intrinsic space one has strings and membranes.   
The {\bf second} is by grading the action with fermionic charges.   
The {\bf third} is to produce deviation equations.   Such 
generalizations can be done in combination and in principle the deviation 
method used here can be applied to the actions corresponding to other 
generalizations, again these points are left for future investigation.
A point particle which carries charge 
Roberts (1989) \cite{bi:mdr2} cannot be 
constructed from a Lagrangian theory because none of
the Lorentz-Dirac,  Hobbs,  and DeWitt-Brehme terms cannot be recovered,
resulting in no Lagrange method being useable here.
Perhaps this is because only {\it part} of the physical system 
is being described,   quantum electrodynamics describes the {\it full}
system and this is Lagrangian based.
The conventions used are the same as in Hawking and Ellis (1973)\cite{bi:HE}
(in particular the signature is $-,+,+,+$),
additionally $\dot{x}^{2}=\dot{x}\cdot\dot{x}=\dot{x}^{\al}\dot{x}_{\al}$.
\section{The First Covariant Variation.}
\subsection{The Point Particle}
The action of a point particle in coordinate space 
is taken to be
\be
S=\int_{\ta_{1}}^{\ta_{2}}d\ta \Lg.
\label{eq:ptpac}
\ee
The standard point particle Lagrangian $\Lg$ is given by
\be
\Lg = - m \times \el^{n},
\label{eq:ptplag}
\ee
where the ``lenght'' $\el$ is
\be
\el \equiv\sqrt{-\dot{x}\cdot\dot{x}}
=\sq{-g_{\al\bt}\fr{Dx^\al}{d\ta}\fr{Dx^\bt}{d\ta}},
\label{eq:ptplenght}
\ee
and $\ta$ is the evolution parameter,  m is the mass,  and the velocity 
is $\dot{x}^{\al}=Dx^{\al}/d{\ta}$.
The action with $n=1$ is reparameterization invariant,  
other values of $n$ represent gauged-fixed actions and are not fuly equivalent
(e.g.,  they have different interpretations in the quantum theory of 
relativistic particles).   Here the $n=1$ reparameterization invariant 
action is used.   Varying the velocity one can
interchange the variation and the covariant derivative thus
\be
\de \dot{x}^{\al}=\de \fr{Dx^{\al}}{d \ta}
=\fr{D\de x^{\al}}{d\ta},
\label{eq:inter1}
\ee
c.f. \cite{bi:bazanski} equation(1.4).   Now varying the action
\be
\de S= \fr{\p \Lg }{\p \dot{x}^{\mu}}
\de x^{\mu}\mid_{\ta_{1}}^{\ta_{2}} 
- \int_{\ta_{1}}^{\ta_{2}}d \ta \de x^{\mu} 
\fr{D}{d\ta}\fr{\p \Lg}{\p \dot{x}^{\mu}}.
\label{eq:ptpva}
\ee
The first term of the action can be made to vanish by choosing the boundary
condition
\be
\de x^{\mu}\fr{\p \Lg}{\p \dot{x}^{\mu}}=0,
\label{eq:bnd1}
\ee
at $\ta_{1}$ and $\ta_{2}$,  or more simply
\be
\de x^{\mu} \mid_{\ta_{1}}=\de x^{\mu} \mid_{\ta_{2}}=0,
\label{eq:bnd2}
\ee
The second term of \ref{eq:ptpva} vanishes when the equation of motion
\be
\fr{D}{d\ta}\fr{\p \Lg}{\p \dot{x}^{\mu}}=0,
\label{eq:ptpeqm}
\ee
is obeyed.   Specifically for the standard Lagrangian \ref{eq:ptplag} 
\be
\fr{\p \Lg}{\p \dot{x}^{\mu}}
=\fr{m\dot{x}^{\mu}}\el,
\label{eq:varptp}
\ee
so that the equation of motion becomes the geodesic equation
\be
\fr{D}{d\ta}\fr{\dot{x}^{\mu}}{\el}=0.
\label{eq:geodesiceq}
\ee
The momentum is defined by
\be
P_{\mu}\equiv\fr{\de S}{\de \dot{x}^{\mu}}=\fr{\p \Lg}{\p \dot{x}^{\mu}},
\label{eq:ptpmom}
\ee
the equality holding because the Lagrangians considered here are both
explicitly defined and are only functions of $\dot{x}^{\mu}$ 
not $\ddot{x}^{\mu}$.
The derivative of the momenta is defined by
\be
\dot{P}^{\mu}\equiv\fr{D}{d \ta}P^{\mu}.
\ee
In terms of the momentum the boundary condition \ref{eq:bnd1} is
\be
\de x^{\mu}P_{\mu}=0,
\ee
and the equation of motion \ref{eq:ptpeqm} is
\be
\dot{P}^{\mu}=0.
\ee
Here througout momenta are introduced last so as to emphasise that the actions
considered are coordinate space actions rather than phase space actions.
\subsection{The String}
The action of a string \cite{bi:scherk} eq.I.16 is
\be
S=\int_{\ta_{1}}^{\ta_{2}}d\ta\int_{0}^{\pi}d\si\Lg,
\label{eq:stringaction}
\ee
where the standard Lagrangian $\Lg$ is given by
\be
\Lg=-\fr{\ar}{2 \pi \al'}
\label{eq:stringlag}
\ee
and the ``area'' $\ar$ is
\ber
\ar&\equiv&\sq{(\dot{x}\cdot x')^{2}-\dot{x}^{2}x'^{2}}\nonumber\\
        &=&\sq{g_{\al\bt}\fr{Dx^\al}{d\ta}\fr{Dx^\bt}{d\si}
               g_{\ga\de}\fr{Dx^\ga}{d\ta}\fr{Dx^\de}{d\si}
              -g_{\al\bt}\fr{Dx^\al}{d\ta}\fr{Dx^\bt}{d\ta}
               g_{\ga\de}\fr{Dx^\ga}{d\si}\fr{Dx^\de}{d\si}},
\label{eq:stringarea}
\ear
and $\ta$ is the evolution parameter, $\si$ is the kinematic parameter,
$\bf{\al '}$ is the string tension,  
$\dot{x}^{\al}\equiv\fr{D x^{\al}}{d\ta},\;
x'^{\al}\equiv\fr{Dx'}{d\si}$,  and the absolute derivatives are 
$\fr{D}{d\ta}\equiv\dot{x}^{\al}\na_{\al},
\fr{D}{d\si}\equiv x'^{\al}\na_{\al}$.
Varying the velocities in a similar manner to \ref{eq:inter1} gives
\ber
\de \dot{x}^{\al}=\fr{D}{d\ta}\de x^{\al},\nonumber\\
\de x'^{\al}=\fr{D}{d\si}\de x^{\al}.
\label{eq:inter2}
\ear
Now varying the action c.f. \cite{bi:scherk} eqs.I.17-19
\ber
\de S=&-&\int_{\ta_{1}}^{\ta_{2}}d\ta
\int_{0}^{\pi} d \si \de x^{\mu}\left(
\fr{D}{d\ta}
\fr{\p \Lg}{\p \dot{x}^{\mu}}
+\fr {D}{d\si}\fr{\p \Lg}{\p x'^{\mu}}
\right)\nonumber\\
&+&\int_{0}^{\pi}d\si \fr{\p \Lg}
{\p \dot{x}^{\mu}}\de x^{\mu}
\mid_{\ta =\ta_{1}}^{\ta =\ta_{2}}\nonumber\\
&+&\int_{\ta_{1}}^{\ta_{2}}d\ta \fr{\p \Lg}{\p x'^{\mu}}
\de x^{\mu}\mid_{\si =0}^{\si = \pi}.
\label{eq:stringva}
\ear
Choosing initial and final positions on the string to be fixed so that 
\ref{eq:bnd2} is obeyed the second term vanishes.   The third term vanishes 
when the edge condition,  c.f. Scherk (1975) \cite{bi:scherk} eq.I.18.1
\be
\fr{\p \Lg}{\p x'^{\mu}}\mid_{\si=0}
=\fr{\p \Lg}{\p x'^{\mu}}\mid_{\si=\pi},
\label{eq:edgecondition}
\ee
is obeyed.   This edge condition is for open strings,
for closed string $x_{\mu}(\ta,\si+2\pi)=x_{\mu}(\ta,\si)$,
c.f. Scherk (1975) \cite{bi:scherk} section II.7.
The vanishing of the first term gives the equation of motion
\be
\fr{D}{d\ta}\fr{\p \Lg}{\p \dot{x}^{\mu}}
+\fr{D}{d\si}\fr{\p \Lg}{\p x'^{\mu}}=0.
\label{eq:stringeqmotion}
\ee
Using the reduction equation
\be
\fr{\p \Lg}{\p x'^{\al}}=0,
\label{eq:re1}
\ee
the string equation of motion \ref{eq:stringeqmotion} 
takes the same form as point particle equation of motion \ref{eq:ptpeqm}.
The reduction equations
\be
\dot{x}\cdot x'=0,\;
x'^{2}=+1,\;
\label{eq:re23}
\ee
reduce the ``area'' \ref{eq:stringarea}
to the ``lenght'' \ref{eq:ptplenght}.
The reduction equation
\be
m=(2\pi \bf{\al '})^{-1},
\label{eq:re4}
\ee
equates the coupling constants so that 
the point particle equation of motion \ref{eq:ptpeqm} is recovered.   
Again one can define momenta
\ber
&P_{\ta\mu}&\equiv\fr{\de S}{\de \dot{x}^{\mu}}
             =\fr{\p \Lg}{\p \dot{x}^{\mu}},\nonumber\\
&P_{\si\mu}&\equiv\fr{\de S}{\de x'^{\mu}}
             =\fr{\p \Lg}{\p x'^{\mu}}.
\label{eq:stringmom}
\ear
Derivatives of the momenta are defined by 
\ber
\dot{P}_{\ta}^{\mu}\equiv\fr{D}{d \ta}P_{\ta}^{\mu},\nonumber\\
P'^{\mu}_{\si}\equiv\fr{D}{d \si}P_{\si}^{\mu}.
\ear
These momenta allow the edge condition \ref{eq:edgecondition}
to be put in the form
\be
P_{\si}^{\mu}\mid_{\si=0}=P_{\si}^{\mu}\mid_{\si=\pi},
\ee 
and equation of motion takes the simple form
\be
\dot{P}_{\ta}^{\mu}+P'^{\mu}_{\si}=0.
\ee
For the specific Lagrangian \ref{eq:stringlag}
the momenta \ref{eq:stringmom} are
\ber
&P_{\ta}^{\mu}&=\fr{-1}{2 \pi \al' \ar}
\left((\dot{x}\cdot x')x'^{\mu}-x'^{2}\dot{x}^{\mu}\right),\nonumber\\
&P_{\si}^{\mu}&=\fr{-1}{2 \pi \al' \ar}
\left((\dot{x}\cdot x')\dot{x}^{\mu}-\dot{x}^{2}x'^{\mu}\right).
\ear
The first reduction equation \ref{eq:re1} becomes
\be
\dot{x}^{2}x'^{\mu}=0
\label{eq:re5}
\ee
\section{The Second Covariant Variation.}
\subsection{The Point Particle Again}
The second variation of the point particle action is
\ber
\de^{2} S
&=&\int_{\ta_{1}}^{\ta_{2}}d\ta \de(\de \Lg)\nonumber\\
&=&\int_{\ta_{1}}^{\ta_{2}}d\ta \de
(\fr{\p \Lg}{\p \dot{x}^{\mu}}\de \dot{x}^{\mu})\nonumber\\
&=&\int_{\ta_{1}}^{\ta_{2}}d\tau\left(
\fr{\p \Lg}
{\p \dot{x}^{\mu}}\de^{2}\dot{x}^{\mu}
-\fr{\p^{2}\Lg}
{\p\dot{x}^{\nu}\p \dot{x}^{\mu}}
\de \dot{x}^{\mu}\de \dot{x}^{\nu}\right).
\label{eq:2ndvarptp}
\ear
The Ricci identity for $\dot{x}^{\alpha}$ is
\be
\de^{2}\dot{x}^{\al}=\fr{D}{D\ta}\de^{2}x^{\al}
+R^{\al}_{. \mu \nu \rh}\de x^{\mu}\de x^{\nu}\dot{x}^{\rh},
\label{eq:ric1}
\ee
c.f. \cite{bi:bazanski} eq.(2.2).   Inserting
\ref{eq:ric1} into \ref{eq:2ndvarptp} gives 
the second covariant variation
\ber
\de^{2}S
&=
-\int_{\ta_{1}}^{\ta_{2}}d \ta \de x ^{\nu}
\left[
\fr{D}{d\ta}(\fr{\p^{2}\Lg}
{\p \dot{x}^{\nu} \p \dot{x}^{\mu}}
\fr{D}{d\ta}\de x^{\mu})
-\fr{\p \Lg}{\p \dot{x}^{\mu}}R^{\mu}_{. \al \nu \rh}
\de x^{\al}\dot{x}^{\al}
\right]\nonumber\\
&-\int_{\ta_{1}}^{\ta_{2}}
d\ta \de^{2}x^{\mu}
\fr{D}{d\ta}\fr{\p \Lg}{\p \dot{x}^{\mu}}\nonumber\\
&\left(
\de^{2}x^{\mu}\fr{\p \Lg}{\p \dot{x}^{\mu}}
+
\fr{\p^{2}\Lg}{\p \dot{x}^{\nu} \p \dot{x}^{\mu}}
\de \fr{D}{d\ta}(\de x^{\mu})\de x^{\nu}
\right)\mid_{\ta=\ta_{1}}^{\ta=\ta_{2}}.
\label{eq:2ndcovarptp}
\ear
The last term vanishes by the first order boundary condition
\ref{eq:bnd2}.   The second to last term vanishes by the second 
order boundary condition
\be
\de^{2} x^{\mu}\mid_{\ta_{2}}=\de^{2}x^{\mu}\mid_{\ta_{1}}=0,
\label{eq:bnd3}
\ee
alternatively it vanishes by the equation of motion \ref{eq:ptpeqm}.
There remains the first integral term.   
{\bf If} the infinitesimal $\de x^{\al}$ inside 
the square bracket is {\bf identified} with the separation vector $r^{\al}$;  
then \ref{eq:2ndcovarptp} vanishes when the deviation equation
\be
\fr{\p \Lg}{\p \dot{x}^{\mu}}
R^{\mu}_{. \al \nu \bt}r^{\al}\dot{x}^{\bt}   
-\fr{D}{d\ta}\left(
\fr{\p^{2}\Lg}{\p \dot{x}^{\nu}\p \dot{x}^{\mu}}
\fr{D}{d\ta}r^{\mu}
\right)=0,
\label{eq:ptpdv}
\ee
is obeyed.   Introducing the momentum \ref{eq:ptpmom}
and the general projection
\be
H_{\mu \nu}\equiv\frac{\p^{2} \Lg}{\p \dot{x}^{\nu} \p \dot{x}^{\mu}}
\label{eq:genproj}
\ee
the point particle deviation equation is
\be
P^{\mu}r^{\al}\dot{x}^{\bt}R_{\mu \al \nu \bt}
-\fr{D}{d \ta}\left(H_{\mu}^{\nu}\dot{r}^{\mu}\right)=0,
\label{eq:ptpde}
\ee
where $\dot{r}^\mu$ is defined by
\be
\dot{r}^{\mu}\equiv\fr{D}{d \ta}r^{\mu}.
\ee
The specific Lagrangian 
\ref{eq:ptplag} has first order variation \ref{eq:varptp} and
the second order variation gives the general projection \ref{eq:genproj}
explicitly as
\be
H^{\mu \nu}=\fr{m}{\el}h^{\mu \nu},
\label{eq:varptp2}
\ee
where the standard projection tensor $h^{\mu \nu}$ is defined by
\be
h^{\mu\nu}\equiv g^{\mu\nu}+\fr{1}{\el^{2}}\dot{x}^{\mu}\dot{x}^{\nu},
\label{eq:proj}
\ee
and has trace $h^{\mu}_{\mu}=d-1$, and d is the dimension of the spacetime.   
Substituting for $P^{\mu}$ using equation  \ref{eq:ptpmom} 
and $H^{\mu \nu}$ using equation \ref{eq:varptp2},
multiplying by $m\el$ 
and then using the algebraic properties of the Riemann tensor 
gives the geodesic deviation equation
\be
R^{\mu}_{\al \bt \ga}\dot{x}^{\al}\dot{x}^{\ga}r^{\bt}
+\el\fr{D}{d\ta}\left(
\fr{1}{\el}
h^{\mu}_{\al}
\dot{r}^{\al}\right)=0.
\label{eq:geodev}
\ee
\newpage
\subsection{The String Again}
For the string action the second variation gives six terms
\ber
\de^{2}S\nonumber\\
&=&\int_{\ta_{1}}^{\ta_{2}}d\ta\int_{0}^{\pi}d\si \de
[\fr{\p \Lg}{\p \dot{x}^{\mu}}\p \dot{x}^{\mu}
+\fr{\p \Lg}{\p x'^{\mu}}\de x'^{\mu}]\nonumber\\
&=&\int_{\ta_{1}}^{\ta_{2}}d\ta\int_{0}^{\pi}d\si [
\fr{\p \Lg}{\p \dot{x}^{\mu}}\de^{2}\dot{x}^{\mu}
+\fr{\p \Lg}{\p x'^{\mu}}\de^{2}x'^{\mu}\nonumber\\
&+&\de
(\fr{\p \Lg}{\p \dot{x}^{\mu}})
\de \dot{x}^{\mu}
+\de(\fr{\p \Lg}{\p x'^{\mu}})\de x'^{\mu}]\nonumber\\
&=&\int_{\ta_{1}}^{\ta_{2}}d\ta\int_{0}^{\pi}d\si
[
\fr{\p \Lg}{\p \dot{x}^{\mu}}\de^{2}\dot{x}^{\mu}
+\fr{\p \Lg}{\p x'^{\mu}}\de^{2}x'^{\mu}\nonumber\\
&+&\fr {\p^{2}\Lg} {\p \dot{x}^{\nu}\p \dot{x}^{\mu}}
 \de \dot{x}^{\mu}\de \dot{x}^{\nu}
+\fr {\p^{2}\Lg} {\p x'^{\nu}\p \dot{x}^{\nu}}
 \de \dot{x}^{\mu}\de x'^{\nu}\nonumber\\
&+&\fr {\p^{2}\Lg} {\p \dot{x}^{\nu}\p x'^{\mu}}
 \de x'^{\mu}\de \dot{x}^{\nu}
+\fr {\p^{2}\Lg} {\p x'^{\nu}\p x'^{\mu}}
 \de x'^{\mu}\de x'^{\nu}].
\label{eq:2ndvarstring}
\ear
In addition to the Ricci identity \ref{eq:ric1} for $\dot{x}^{\al}$ 
use the Ricci identity for $x'^{\al}$
\be
\de^{2}x'^{\al}=\fr{D}{d\si}\de^{2}x^{\al}
+R^{\al}_{. \mu \nu \rh}\de x^{\mu}x^{\nu}x'^{\rh},
\label{eq:ric2}
\ee
to give the twice varied action
\ber
\de^{2}S
&=&\int_{\ta_{1}}^{\ta_{2}}d\ta\int_{0}^{\pi}d\si
[
\fr{\p \Lg}{\p \dot{x}^{\al}}
\fr{D}{d\ta}\de^{2}x^{\al}+\fr{\p \Lg}{\p \dot{x}^{\al}}
R^{\al}_{. \mu \nu \rh}\de x^{\mu}x^{\nu}\dot{x}^{\rh}\nonumber\\
&+&\fr{\p \Lg}{\p x'^{\al}}
\fr{D}{d\si}\de^{2}x^{\al}
+\fr{\p \Lg}{\p x'^{\al}}
R^{\al}_{. \mu \nu \rh}\de x^{\mu}\de x^{\nu}x'^{\rh}\nonumber\\
&+&\fr{\p^{2}\Lg}{\p \dot{x}^{\nu}\p \dot{x}^{\mu}}
\de \dot{x}^{\mu}\fr{D}{d\ta}\de x^{\nu}
+\fr{\p^{2}\Lg}{\p x'^{\nu}\p \dot{x}^{\mu}}
\de \dot{x}^{\mu}\fr{D}{d\si}\de x^{\nu}\nonumber\\
&+&\fr{\p^{2}\Lg}{\p \dot{x}^{\nu}\p x'^{\nu}}
\de x'^{\nu}\fr{D}{d\ta}\de x^{\nu}
+\fr{\p^{2}\Lg}{\p x'^{\nu}\p x'^{\mu}}
\de x'^{\mu}\fr{D}{d\si}\de x^{\nu}].
\label{eq:311}
\ear
The first and third terms vanish if the first order equations of motion are 
assumed and also the second order boundary conditions
\be
\de^{2}x^{\al}\mid_{\ta_{1}}
=\de^{2}x^{\al}\mid_{\ta_{2}}
=\de^{2}x^{\al}\mid_{\si =0}
=\de^{2}x^{\al}\mid_{\si = \pi}=0.
\label{eq:bnd4}
\ee
leaving
\ber
\de^{2}S
=\int_{\ta_{1}}^{\ta_{2}}d\ta\int_{0}^{\pi}d\si
(
\fr{\p\Lg}{\p \dot{x}^{\al}}\de \dot{x}^{\al}
+\fr{\p \Lg}{\p x'^{\al}}\de x'^{\al})
\de x^{\mu} \de x^{\nu} R^{\al}_{.\mu \nu \rh}\nonumber\\
+[\int_{0}^{\pi}d\si(
\fr{\p^{2}\Lg}{\p \dot{x}^{\nu}\p \dot{x}^{\mu}}
\de \dot{x}^{\mu}
+\fr{\p^{2} \Lg}{\p \dot{x}^{\nu}\p x'^{\mu}}\de x'^{\mu}
)\de x^{\nu}]\mid_{\ta_{1}}^{\ta_{2}}\nonumber\\
-\int_{\ta_{1}}^{\ta_{2}}d\ta\int_{0}^{\pi}d\si
\de x^{\nu}\fr{D}{d\ta}(
\fr{\p^{2}\Lg}{\p \dot{x}^{\nu}\p \dot{x}^{\mu}}
\de \dot{x}^{\mu}
+\fr{\p^{2}\Lg}{\p \dot{x}^{\nu}\p x'^{\mu}}\de x'^{\mu} )
\nonumber\\
+[\int_{\ta_{1}}^{\ta_{2}}d\ta(
\fr{\p^{2}\Lg}{\p x'^{\nu}\p \dot{x}^{\mu}}
\de \dot{x}^{ \mu}
+\fr{\p^{2}\Lg}{\p x'^{\nu}x'^{\mu}}\p x'^{\mu})
\de x'^{\nu}]\mid_{\si=0}^{\si = \pi}\nonumber\\
-\int_{\ta_{1}}^{\ta_{2}}d\tau\int_{0}^{\pi}d\si
\de x^{\nu}\fr{D}{d\si}(
\fr{\p^{2}\Lg}{\p x'^{\nu}\delta \dot{x}^{\mu}}
\de \dot{x}^{\mu}
+\fr{\p^{2}\Lg}{\p x'^{\nu}\p x'^{\mu}}\de x'^{\mu}).
\label{eq:313}
\ear
The second to last term vanishes by the edge condition \ref{eq:edgecondition}
of the first variation.   The second term vanishes by the condition 
\ref{eq:bnd2} leaving
\ber
\de^{2}S&=&
\int_{\ta_{1}}^{\ta_{2}}\int_{0}^{\pi}d\si\de x^{\nu}[\left(
\fr{\p \Lg}{\p \dot{x}^{\al}}\de \dot{x}^{\bt}
+\fr{\p \Lg}{\p x'^{\al}}\de x'^{\bt}\right)
\de x^{\mu}R^{\al}_{. \mu \nu \rh}\nonumber\\
&-&\fr{D}{d\ta}(
\fr{\p^{2}\Lg}{\p \dot{x}^{\nu}\p \dot{x}^{\mu}}
\de \dot{x}^{\mu}
+\fr{\p^{2}\Lg}{\p \dot{x}^{\nu}\p x'^{\mu}}
\de x'^{\mu})\nonumber\\
&-&\fr{D}{d\si}\left(
\fr{\p^{2}\Lg}{\p x'^{\nu} \p \dot{x}^{\mu}}
\de \dot{x}^{\mu}
+\fr{\p^{2}\Lg}{\p x'^{\nu} \p x'^{\mu}}
\de x'^{\mu}\right)].
\label{eq:314}
\ear
Now $\de x^\mu$ is {\bf identified} with the separation vector $r^\mu$.
No rigorous proof that the indentification holds by necessity is given here.
There are three justifications for this identification.
{\it Firstly},  this identification is {\sc analogous} 
to that of the geodesic case.
{\it Secondly},  there is no {\sc alternative} vector which could be used 
except ones differentiated with respect to either $\ta$ or $\si$.
{\it Thirdly},  {\sc post-hoc} it produces simple equations.
\ber
\left(\fr{\p \Lg}{\p \dot{x}^{\al}}\dot{x}^{\bt}
+\fr{\p \Lg}{\p x'^{\al}}\de x'^{\bt}\right)
r^{\mu}R^{\al}_{.\mu \nu \bt}\nonumber\\
-\fr{D}{d\ta}\left(\fr{\p^{2}\Lg}{\p \dot{x}^{\nu}\p \dot{x}^{\mu}}
\fr{D}{d\ta}r^{\mu}
+\fr{\p^{2}\Lg}{\p \dot{x}^{\nu}\p x'^{\mu}}
\fr{D}{d\si}r^{\mu}\right)\nonumber\\
-\fr{D}{d\si}\left(
\fr{\p^{2}\Lg}{\p x'^{\nu}\p \dot{x}^{\mu}}
\fr{D}{d\ta}r^{\mu}
+\fr{\p^{2}\Lg}{\p x'^{\nu}\p x'^{\mu}}
\fr{D}{d\si}r^{\mu}\right)=0.
\label{eq:sde}
\ear
Using the first reduction equation \ref{eq:re1}
and also the second order reduction equations
\be
\fr{\p^{2} \Lg}{\p \dot{x}^{\nu} \p x'^{\mu}}=
\fr{\p^{2} \Lg}{\p x'^{\nu} \p \dot{x}^{\mu}}=0,
\label{eq:re6}
\ee
and
\be
\fr{\p^{2} \Lg}{\p x'^{\nu} \p x'^{\mu}}=0,
\label{eq:re7}
\ee 
the deviation equation \ref{eq:ptpdv} is recovered.
Define the general projecta
\ber
H_{\ta \ta \mu \nu}
&\equiv&\fr{\p^{2} \Lg}{\p \dot{x}^{\mu} \p \dot{x}^{\nu}},\nonumber\\
H_{\ta \si \mu \nu}
&\equiv&\fr{\p^{2} \Lg}{\p \dot{x}^{\mu} \p x'^{\nu}},\nonumber\\
H_{\si \ta \mu \nu}
&\equiv&\fr{\p^{2} \Lg}{\p x'^{\mu} \p x^{\nu}},\nonumber\\
H_{\si \si \mu \nu}
&\equiv&\fr{\p^{2} \Lg}{\p x'^{\mu} \p x'^{\nu}},
\label{eq:genstringproj}
\ear
note that $H_{\ta \si}^{\mu \nu}$ does not equal $H_{\si \ta}^{\mu \nu}$
unless the matrix is symmetric.   The four projecta are best expressed 
in terms of their symmetric and antisymmetric parts,  thus:
$H_{\ta \ta}^{\mu \nu}$,  $H_{\si \si}^{\mu \nu}$,  
$H_{\ta \si}^{(\mu \nu)}$,  and $H_{\ta \si}^{[\mu \nu]}$.
Using the momenta \ref{eq:stringmom}, 
the string deviation equation is
\ber
\left(P_{\ta}^{\al}\dot{x}^{\bt}+
      P_{\si}^{\al}x'^{\bt}\right)
r^{\mu}R_{\al \mu \nu \bt}~~~~~~~~~~~~~~~~~~~~~~~~~~~~~~~~\nonumber\\
-\fr{D}{d \ta}\left(H_{\ta \ta}^{\nu \mu}\dot{r}_{\mu}
                  +H_{\ta \si}^{\nu \mu}r'_{\mu}\right)
-\fr{D}{d \si}\left(H_{\si \ta}^{\nu \mu}\dot{r}_{\mu}
                  +H_{\si \si}^{\nu \mu}r'_{\mu}\right)=0.
\label{eq:stringdevH}
\ear
For the standard Lagrangian \ref{eq:stringlag}
the projection tensors become
\ber
H_{\ta \ta}^{\mu \nu}
&=&\fr{x'^{2}}{2 \pi \al' \ar}h^{\mu \nu},\nonumber\\
H_{\si \si}^{\mu \nu}
&=&\fr{\dot{x}^{2}}{2 \pi \al' \ar}h^{\mu \nu},\nonumber\\
H_{\ta \si}^{(\mu \nu)}
&=&\fr{-(\dot{x} \cdot x')h^{\mu \nu}}{2 \pi \al' \ar}\nonumber\\
H_{\si \ta}^{[\mu \nu]}
&=&\fr{1}{2 \pi \al' \ar}
     \left(-\dot{x}^{\mu}x'^{\nu}+x'^{\mu}\dot{x}^{\nu}\right),
\ear
where the standard projection tensor is defined by
\be
h^{\mu \nu}\equiv g^{\mu \nu}+\fr{1}{\ar^{2}}
(\dot{x}^{2}x'^{\mu}x'^{\nu}
-(\dot{x}\cdot x')(\dot{x}^{\mu}x'^{\nu}+x'^{\mu}\dot{x}^{\nu})
+x'^{2}\dot{x}^{\mu}\dot{x}^{\nu}),
\ee
and has trace $h^{\mu}_{\mu}=d-2$.
Using the reduction equations \ref{eq:re23},\ref{eq:re5}
this projection tensor reduces to \ref{eq:proj}.
In the specific case of the standard Lagrangian \ref{eq:stringlag},
multipying by $-2 \pi \al' \ar $ and substituting gives the standard string
deviation
\ber
\ar(g^{\al \bt}-h^{\al \bt})r^{\mu}
        R_{\al \mu \nu \bt}~~~~~~~~~~~~~~~~~~~~~~~~~~~~\nonumber\\
-\fr{D}{d \ta}\fr{1}{\ar}\left(h_{\mu}^{\nu}
     (x'^{2}\dot{r}^{\mu}-\dot{x}\cdot x' r'^{\mu})+r'^{\mu}
(\dot{x}^{\mu}x'^{\nu}-x'^{\mu}\dot{x}^{\nu})\right)\nonumber\\
-\fr{D}{d \si}\fr{1}{\ar}\left(h^{\nu}_{\mu}
     (-\dot{x}\cdot x' \dot{r}^{\mu}+\dot{x}^{2}r'^{\mu})
   +\dot{r}^{\mu}(-\dot{x}^{\mu} x'^{\nu}
   +x'^{\mu}\dot{x}^{\nu})\right)=0.
\label{eq:stanstringdev}
\ear
For this case the reduction equation \ref{eq:re6} is satisfied identically
and \ref{eq:re7} becomes
\be
\fr{D}{d \si}\fr{1}{\ar}h^{\nu}_{\mu}\dot{x}^{2}r'^{\mu}=0.
\label{eq:re8}
\ee
Applying the reduction equations \ref{eq:re23},\ref{eq:re5}, and \ref{eq:re8}
to the standard string deviation \ref{eq:stanstringdev}
gives the geodesic deviation equation \ref{eq:geodev}.
\section{The Combined Action.}\label{sec:combined}
\subsection{The Point Particle Yet Again}
The equations produced from a second covariant variation can often
be derived from a combined action.   For the point particle consider
\be
S_{c}=\int_{\ta_{1}}^{\ta_{2}}d\ta
\fr{\p \Lg}{\p \dot{x}^{\al}}
\fr{D}{d\ta}r^{\al},
\label{eq:ptpca}
\ee
with $\ta$ the evolution parameter and $r^{\al}$  a separation vector.
The Ricci identity can be expressed as
\be
\de\left(\fr{D}{d\ta}r^{\al}\right)
=\fr{D}{d\ta}\de r^{\al}
+R^{\al}_{. \la \mu \nu}r^{\la}\de x^{\mu}\dot{x}^{\nu},
\label{eq:ric3}
\ee
Varying the action gives
\ber
\de S_{c}&=&\int_{\ta_{1}}^{\ta_{2}}d\ta[
\de\left(\fr{\p \Lg}{\p \dot{x}^{\al}}\right)
\fr{D}{d\ta}r^{\al}
+\fr{\p \Lg}{\p \dot{x}^{\al}}
R^{\al}_{. \la \mu \nu}r^{\la}\de x^{\mu}\dot{x}^{\nu}]\nonumber\\
&+&[\fr{\p \Lg}{\p \dot{x}^{\al}}\de r^{\al}]\mid_{\ta_{1}}^{\ta_{2}}
-\int_{\ta_{1}}^{\ta_{2}}d\ta \de r^{\al}
\fr{D}{d\ta}\fr{\p \Lg}{\p \dot{x}^{\al}},
\label{eq:ptpcava}
\ear
Assuming that $\Lg$ is only an explicit function of $\dot{x}^{\al}$
and not of $x^{\al},r^{\al},\dot{r}^{\al}$ gives
\ber
\delta S_{c}
&=&\int_{\ta_{1}}^{\ta_{2}}d \ta [
-\fr{D}{d\ta}(
\fr{\p^{2}\Lg}{\p \dot{x}^{\bt}\p \dot{x}^{\al}}\fr{D}{d\ta}r^{\al})
+\fr{\p \Lg}{\p \dot{x}^{\al}}
R^{\al}_{. \la \bt \nu}r^{\la}\dot{x}^{\nu})]\de x^{\bt}\nonumber\\
&-&\int_{\ta_{1}}^{\ta_{2}}d\ta \de r
\fr{D}{d\ta}\fr{\p \Lg}{\p \dot{x}^{\al}}\nonumber\\
&+&[\de x^{\bt}
\fr{\p^{2}\Lg}{\p \dot{x}^{\bt} \p \dot{x}^{\al}}
\fr{D}{d\ta}r^{\al}
+\fr{\p \Lg}{\p \dot{x}^{\al}}
\de r^{\al}]\mid_{\ta_{1}}^{\ta_{2}}.
\label{eq:44}
\ear
The boundary term vanishes if
\be
\de x^{\al}\mid_{\ta_{1}}
=\de x^{\al}\mid_{\ta_{2}}
=\de r^{\al}\mid_{\ta_{1}}
=\de r^{\al}\mid_{\ta_{2}}=0.
\label{eq:bnd5}
\ee
The $\de r^{\al}$ integral term vanishes if the particle equation 
of motion \ref{eq:ptpeqm} is obeyed,  the $\de x^{\al}$  
integral term vanishes if the particle deviation equation
\ref{eq:ptpdv} is obeyed.   Defining momenta
\ber
P^{\mu}\equiv\fr{\de S_{c}}{\de \dot{r}^{\mu}}
&=&\fr{\p \Lg}{\p \dot{x}^{\mu}},\nonumber\\
\Pi^{\mu}\equiv\fr{\de S_{c}}{\de \dot{x}^{\mu}}
&=&\dot{r}^{\nu}\fr{\p^{2} \Lg}{\p \dot{x}^{\mu}\p \dot{x}^{\nu}}
=\dot{r}^{\nu}H^{\mu}_{\nu},
\ear
and
\be
\dot{\Pi}^{\mu}\equiv\fr{D}{d \ta}\Pi^{\mu},
\ee
the geodesic deviation equation \ref{eq:ptpde}
takes the simple form
\be
\dot{\Pi}^{\mu}=R^{\mu}_{. \al \bt \ga}r^{\ga}P^{\bt}\dot{x}^{\al}.
\ee
\subsection{The String Yet Again}
For the string consider the combined action
\be
S_{c}=\int_{\ta_{1}}^{\ta_{2}}d\ta \int_{0}^{\pi}d \si \left(
\fr{\p \Lg}{\p \dot{x}^{\al}}
\fr{D}{d\ta}r^{\al}
+\fr{\p \Lg}{\p x'^{\al}}
\fr{D}{d\si}r^{\al}\right)
\label{:stringca}
\ee
Varying
\ber
\de S_{c}=\int_{\tau_{1}}^{\ta_{2}}d \ta \int_{0}^{\pi}d\si
~~~~~~~~~~~~~~~~~~~~~~~~~~~~~~~~~~~~~~~~~~~~~~~~~~~~~~~~~~~\nonumber\\
\vb+{+-\de r^{\al}\left(
\fr{D}{d\ta}\fr{\p \Lg}{\p \dot{x}^{\al}}
+\fr{D}{d\si}\fr{\p \Lg}{\p x'^{\al}}\right)
+\de x ^{\bt}\left(
\dot{x}^{\nu}\fr{\p \Lg}{\p \dot{x}^{\nu}}
+x'^{\nu}\fr{\p \Lg}{\p x'^{\nu}}
\right)R^{\al}_{. \la \bt \nu}r^{\la}\vb+}+\nonumber\\
-\fr {D}{d\ta }\left(
\fr{\p^{2}\Lg}{\p \dot{x}^{\bt}\p \dot{x}^{\al}}
\fr{D}{d\ta}r^{\al}
+\fr{\p^{2}\Lg}{\p \dot{x}^{\bt} \p x'^{\al}}
\fr{D}{d\si}r^{\al}\right)~~~~~~~~~~~~~~~~~\nonumber\\
-\fr{D}{d\si}\left(
\fr{\p^{2}\Lg}{\p x'^{\bt}\p \dot{x}^{\al}}
\fr{D}{d\ta}r^{\al}
+\fr{\p^{2}\Lg}{\p x'^{\bt} \p x'^{\al}}
\fr{D}{d\ta}r^{\al}\right)~~~~~~~~~~~~~~~~~\nonumber\\
+[\int_{0}^{\pi}d\si\left(
\de r^{\al}\fr{\p \Lg}{\p \dot{x}^{\al}}
+\de x^{\bt}(
\fr{\p^{2}\Lg}{\p \dot{x}^{\bt} \p \dot{x}^{\al}}
\fr{D}{d\ta}r^{\al}
+\fr{\p^{2}\Lg}{\p \dot{x}^{\bt}\p x'^{\al}}
\fr{D}{d\si}r^{\al}
)\right)]\mid_{\ta=\ta_{1}}^{\ta=\ta_{2}}\nonumber\\
+[\int_{\ta_{1}}^{\ta_{2}}d\ta\left(
\de r^{\al}\fr{\p \Lg}{\p x'^{\al}}
+\de x^{\bt}(
\fr{\p^{2}\Lg}{\p x'^{\bt}\p \dot{x}^{\al}}
\fr{D}{d\ta}r^{\al}
+\fr{\p^{2}\Lg}{\p x'^{\bt}\p x'^{\al}}
\fr{D}{d\si}r^{\al})\right)]\mid_{\si=0}^{\si=\pi}.
\label{eq:48}
\ear
The first integral term vanishes by the string equation of motion 
\ref{eq:stringeqmotion}.
The second integral term vanishes by the string deviation equation
\ref{eq:sde}.
The boundary terms vanish if \ref{eq:bnd5} is obeyed.
Using the general projecta \ref{eq:genstringproj} and defining
\ber
P_{\ta \mu}\equiv\fr{\de S_{c}}{\de \dot{r}^{\mu}}
&=&\fr{\p \Lg}{\p \dot{x}^{\mu}},\nonumber\\
P_{\si \mu}\equiv\fr{\de S_{c}}{\de r'^{\mu}}
&=&\fr{\p \Lg}{\p x'^{\mu}},\nonumber\\
\Pi_{\ta \mu}\equiv\fr{\de S_{c}}{\de \dot{x}^{\mu}}
&=&\dot{r}_{\al}H_{\ta \ta}^{\mu \nu}+r'_{\al}H_{\ta \si}^{\mu \nu},\nonumber\\
\Pi_{\si \mu}\equiv\fr{\de S_{c}}{\de x'^{\mu}}
&=&\dot{r}_{\al}H_{\si \ta}^{\mu \al}+r'_{\al}H_{\si \si}^{\mu \al},
\ear
and substituting into \ref{eq:stringdevH}
the string deviation equation takes the simple form
\be
\dot{\Pi}_{\ta}^{\mu}+\Pi'^{\mu}_{\si}
=R^{\mu}_{\al \bt \ga}
r^{\ga}(P_{\ta}^{\bt}\dot{x}^{\al}+P_{\si}^{\bt}x'^{\al}).
\ee
\section{Conclusion.}
A point particle obeys the geodesic equation
\ref{eq:ptpeqm} and many particles have relative motion described by
the geodesic deviation equation \ref{eq:ptpdv}.   A single string obeys the 
equation of motion \ref{eq:stringeqmotion} and second order variation 
of the action gives the string deviation equation \ref{eq:sde}.   
For the standard point particle Lagrangian \ref{eq:ptplag}
the geodesic equation is \ref{eq:geodesiceq}
and the geodesic deviation equation is \ref{eq:geodev};
for the standard string Lagrangian \ref{eq:stringlag}
the equation of motion is \ref{eq:stringeqmotion}
and the deviation equation is \ref{eq:stanstringdev}.
The second order variations can be expressed in combined first order form,
this leads to the deviation equations of the abstract.
The string deviation equation is a candidate 
for the description of the relative motion of many strings.   
The combined actions of section \ref{sec:combined}
describe a new gauge theory to which
the full paraphernalia of gauge theory can be applied.
\section{Acknowledgements}
I would like to thank T.W.B.Kibble for his interest in this work,
and also A.L.Larsen for pointing out similar work on perturbing strings
by himself and Frolov \cite{bi:FL},  
Carter \cite{bi:carter},  and Guven \cite{bi:guven}.
This work has been supported in part by the South African 
National Research Foundation,  formally called the Foundation 
for Research and Development (FRD).

\end{document}